\documentclass[9pt,twocolumn,twoside,dvipsnames,svgnames]{opticajnl}
\journal{optica} % Choose journal (ao,jocn,josaa,josab,ol,optica,pr)

\setboolean{shortarticle}{false}
\usepackage{lineno}
\usepackage{tikz}
\usetikzlibrary{scopes}
\usepackage[subpreambles=true]{standalone}
\usepackage{import}
\usepackage{pgfplots}
\pgfplotsset{compat=1.18}
\newcommand{\figref}[1]{Fig.~\ref{#1}}

\title{\vspace{-2em}Efficient, ultra-high attenuation fiber Bragg grating filter for photon noise suppression}
\usepgfplotslibrary{external} 
\usepackage{lmodern}
\author[1,2,*]{Benjamin R. Field}
\author[1]{Chintan Mistry}
\author[3]{Liguo Luo}
\author[3]{Goran Edvell}
\author[4,5]{Joss Bland-Hawthorn}
\author[4,5]{Sergio Leon-Saval}
\author[1,6,*]{John G. Bartholomew}

\affil[1]{Centre for Engineered Quantum Systems, School of Physics, University of Sydney, NSW 2006, Australia}
\affil[2]{Sydney Quantum Academy, Sydney, New South Wales, Australia.}
\affil[3]{Advanced Fiber Bragg Grating Facility, Research and Prototype Foundry, University of Sydney, NSW 2006, Australia}
\affil[4]{Sydney Astrophotonic Instrumentation Labs, School of Physics, University of Sydney, NSW 2006, Australia}
\affil[5]{Sydney Institute for Astronomy, School of Physics, University of Sydney, NSW 2006, Australia}
\affil[6]{University of Sydney Nano Institute, University of Sydney, NSW 2006, Australia}

\affil[*]{Corresponding authors: benjamin.field@sydney.edu.au, john.bartholomew@sydney.edu.au}

\begin{abstract}
Precision optical filters are key components for current and future photonic technologies. Here, we demonstrate a low loss spectral filter consisting of an ultrasteep bandpass feature with a maximum gradient of (90.6$\pm$0.7)dB/GHz, centred within a notch filter with (128$\pm$6)~dB of suppression. The filter consists of a fiber Bragg grating with multiple $\pi$-phase discontinuities inscribed into a single mode photosensitive fiber. The measured performance closely matches the simulated spectrum calculated from the design parameters indicating a high degree of confidence in the repeatability and manufacture of such devices. These filters show great promise for applications reliant on high-frequency resolution noise suppression, such as quantum networking, and highlight the opportunities for the versatility, efficiency, and extreme suppression offered by high-performance fiber Bragg grating devices.
\end{abstract}

\setboolean{displaycopyright}{false}

\begin{document}

\maketitle

\section{Introduction}

There is an increasing need for the development of versatile, low loss, high-frequency resolution optical filters for advanced photonic technology development. Applications as diverse as photonic computing, telecommunications, medicine, astronomical measurement and sensing\cite{taha_2024,moody_2022} all drive demand for higher performance optical filters. Quantum computing and communication technologies provide even more stringent demands on photonic filter loss and noise specifications, both of which are crucial to scaling systems toward utility\cite{alexander_2024,bourassa_2021},. Irrespective of the number of components in a quantum optics system, the absolute maximum loss budget must be well below 3~dB, and any added noise should be significantly less than a single photon in the optical mode of interest.
%%%%%%%%%%%%%%%%%%%%%%%%%%%%%%%%%%%%%%%%%%%%%%%%%%%%%%%%%%%%%%%%%%%
\begin{figure}[ht!]
\centering
\resizebox{\linewidth}{!}{\import{./figures}{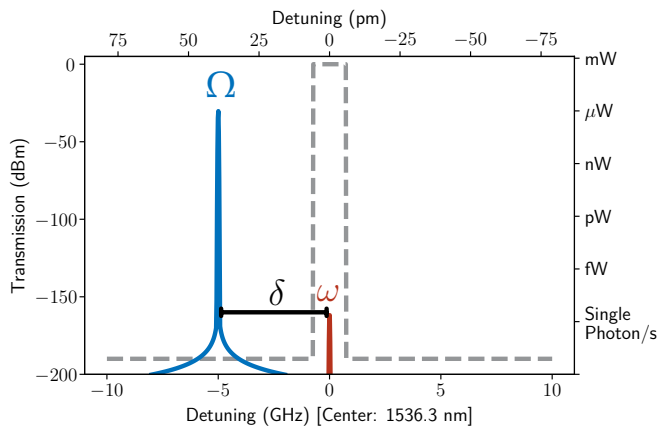}}
\caption{Indicative input classical pump leakage signal (blue) and output quantum signal (red) for a microwave to optical transducer. A classical pump beam $\Omega$ is required to bridge the energy gap between the input microwave photon at frequency $\delta$ and the optical photon $\omega$. The majority of the input pump energy is consumed in the transduction process but a few microWatts of pump power leak into the output spatial mode of interest. To resolve single photon transduction signals from this leakage requires a high performance filter such as the idealizer spectrum indicated by the dashed grey line. In our work the filter central frequency is chosen to be compatible with erbium-based transduction schemes between 5~GHz microwave signals, and the telecom C band~\cite{williamson_2014}.}
% \caption{Example of the common signals resulting from Microwave to optical transduction, and the required filtering. A classical pump beam $\Omega$ is required to bridge the energy gap between the input microwave photon at frequency $\delta$ and the output telecom optical photon $\omega$. For our case the central filter frequency is chosen to be compatible with erbium based transduction schemes. (y-axis not to scale)} 

\label{f:transduction}
\end{figure}
%%%%%%%%%%%%%%%%%%%%%%%%%%%%%%%%%%%%%%%%%%%%%%%%%%%%%%%%%%%%%%%%%%%

In this work, we focus on designing and characterizing an optical filter that supports the creation of a near-infrared photonic link between quantum devices operating in the microwave regime. This link is made possible using a quantum transducer, a device that is able to convert between microwave and optical signals at the quantum level~\cite{lambert_2020, lauk_2020,han_2021}. The ideal quantum transducer could convert a single microwave photon to a single optical photon with 100\% efficiency without adding noise. To drive the nonlinearity that can couple these frequency-disparate modes, the majority of solutions require at least one classical optical pump~\cite{barnett_2020,mirhosseini_2020,williamson_2014}. A filter is required to suppress the classical pump light that leaks into the output optical mode of interest and obscures the upconverted quantum signal, which is frequency shifted by only $\approx$ 5~GHz ($\approx 40$~pm at 1550~nm). Based on current implementations~\cite{mirhosseini_2020,rochman_2023,urmey_2023}, intensity suppression at the pump frequency needs to be greater than 100~dB relative to the signal frequency, which ideally is transmitted through the filter with zero insertion loss. An example of the frequency separations and relative signal strengths is shown along with an ideal filter in \figref{f:transduction}.

Current state of the art narrow-band, high dynamic range optical filtering techniques rely on three main mechanisms: interference, atomic absorption and orthogonal modes. Interference based filters rely on the constructive and destructive interference of light in resonators or other layered dielectric structures. The transmission through fiber-coupled Fabry-Perot resonators can achieve suppression depths up to 80~dB at a 5~GHz detuning from resonance with an insertion loss of approximately 2 \textendash 3~dB~\cite{mirhosseini_2020,rochman_2023,urmey_2023}. Translating this performance for transduction experiments continues to be challenging with concatenated filter systems to achieve the required suppression incurring total insertion loss greater than 10~dB~\cite{mirhosseini_2020,rochman_2023,urmey_2023}. While not yet employed for transduction measurements, on-chip interference filters offer attenuation approaching 100~dB, although at a significantly larger detuning from the pass region (several nanometers) ~\cite{oser_2019,alexander_2024}. Such work demonstrates that it is possible to reduce insertion losses to as low as 1~dB, including coupling light on and off chip~\cite{alexander_2024}. It is important to note that interference filters typically reject light by reflecting it back along its initial path. To prevent reflections impacting the transducer performance or resonances between concatenated filters, it is often necessary to use isolators or circulators. Adding components to the optical path will increase the overall loss by of the order of 1~dB per component. The fiber Bragg grating filters (FBGs) discussed in our work are an example of interference-based filters.

An alternate technology to an interference filter is an absorptive filter that relies on narrow, high optical depth atomic transitions. While not common in transduction experiments, these filters have been explored for similarly demanding applications such as quantum memories~\cite{heifetz_2004,stack_2015_1}, and ultrasound detection~\cite{beavan_2013,Ulrich2022}, which often require pass or blocking bandwidths significantly narrower than 1~GHz. Atomic vapour cell filters have achieved suppression depths up to 30~dB harnessing direct absorption~\cite{heifetz_2004,stack_2015_1} or the Faraday effect~\cite{rotondaro_2015}. Crystals containing rare-earth ions at cryogenic temperatures have also been used to demonstrate narrow absorptive spectral features with greater than 50~dB of suppression~\cite{beavan_2013,Ulrich2022}, with a potential to increase to 160~dB~\cite{beavan_2013}. For atomic filters in general, challenges remain to reduce the insertion loss to be comparable with interference filters, mitigate noise processes from coherent and spontaneous emission, and enable continuous operation. 

It is also possible, in general, to leverage orthogonal modes to filter strong pump noise from weak signals. For example, orthogonal polarizations have been used to increase the suppression depth of atomic vapour filters to 40~dB at which point polarization imperfections become the limit~\cite{stack_2015_1}. Similarly, orthogonal spatial modes are used in processes such as single-photon generation to separate the pump along a different axis of the device~\cite{hennrich_2004,huber_2020}. Such orthogonal mode strategies, however, are generally not suitable for transduction, which is often based on coherent wave-mixing requiring mode overlap between pump and signal. 

In this work, we demonstrate the potential for a {\it single} filter, a fiber Bragg grating with multiple $\pi$-phase discontinuities (a $\pi$FBG), to reach the required performance for quantum transduction experiments, i.e. attenuation sufficient to suppress residual pump signals by $>100$~dB to below the single photon level and insertion loss of the order of 1~dB. The filter operates at room temperature, is easily integrated into any fiber-based setup, and is tuneable through strain and temperature. The measured properties of these $\pi$FBG filters and the potential to optimize and customize their properties highlight their appeal as a critical element for current and future high frequency resolution, low noise photonic technologies.

\section{Design and fabrication of a $\pi$FBG}
\begin{figure}[h!]
\centering

{\resizebox{\linewidth}{!}{\import{./figures}{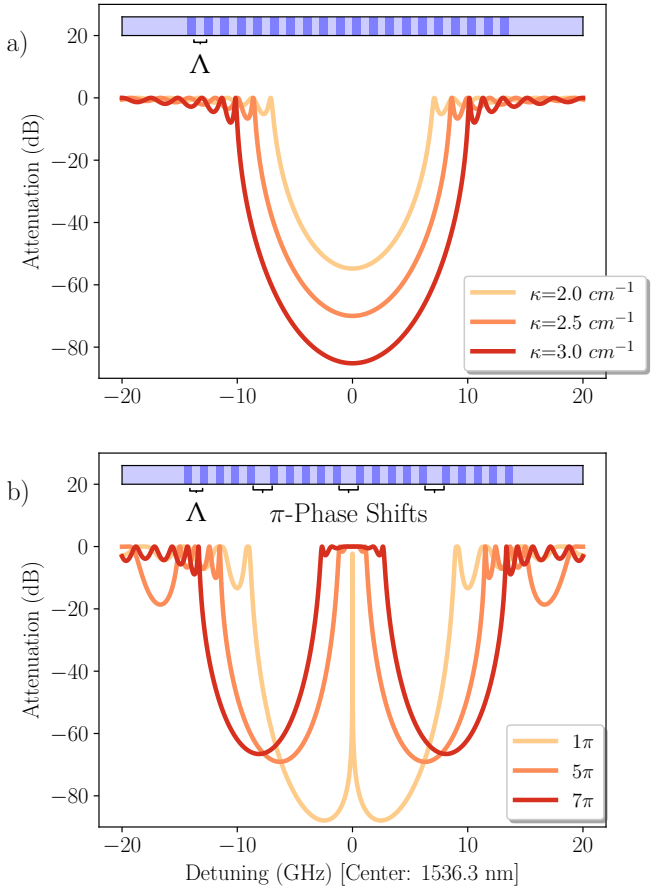}}}
\caption{a) Spectral response of a series of conventional FBG filters, schematically illustrated in the inset. Light blue regions represent the core of the fiber and dark blue regions the UV-induced refractive index changes with period $\Lambda$. Increasing $\kappa$, as defined in \eqref{E:FBG_Kappa}, increases the depth and width of the notch filter. b) Spectral response of a series of phase-shifted FBG filters, schematically illustrated in the inset. The introduction of a $\pi$ phase shift to a conventional FBG creates a narrow pass band region. The pass band width can be increased by increasing the number of $\pi$ phase shifts at the expense of the filter depth. In this example $\kappa = 3$~cm$^{-1}$ is held constant for all three filters.}
\label{f:pi}
\end{figure}

The fiber Bragg grating structure is realised by creating a refractive index modulation in
the core of a single mode fiber~\cite{ghosh_2023,deepa_2020,kori_2020,erdogan_1997}. The period of the modulation $\Lambda$ to achieve the desired center wavelength $\lambda_{\rm eff}$ is set by the Bragg equation 
\begin{equation}
    \lambda_{\rm eff} = 2 n_o \Lambda~,
\end{equation}
where $n_o$ is the mean refractive index of the fiber core. FBGs exploit the weak grating limit (i.e. $\delta n_o/n_o \sim 10^{-4}$) such that a grating can consist of order 100,000 periods over a fiber length of 50~mm. A schematic of such a filter and the resultant spectra as the strength of the refractive index change is increased  is shown in \figref{f:pi} a).

%%%%%%%%%%%%%%%%%%%%%%%%%%%%%%%%%%%%%%%%%%%%%%%%%%%%%%%%%%%%%%%%%%%
%%%%%%%%%%%%%%%%%%%%%%%%%%%%%%%%%%%%%%%%%%%%%%%%%%%%%%%%%%%%%%%%%%%

The $\pi$-phase shifted grating~\cite{agrawal_1994,erdogan_1997} is realised by splitting the conventional FBG into two identical sub-FBGs with a half-period phase difference between them, as shown in \figref{f:pi} b). This can be thought of as a bandwidth-limited Fabry-Perot cavity using two highly reflective mirrors that produce an extremely narrow notch in transmission. By adding more phase shifts, it is possible to broaden the central pass band feature and the FBG notch-band feature at the cost of overall notch depth. The decreased notch depth can be compensated by increasing the effective refractive index modulation ($\bar{\delta n}$), related to the constant $\kappa$ as per~\eqref{E:FBG_Kappa}~\cite{erdogan_1997}. 
\begin{equation}
    \bar{\delta n}=\frac{\lambda}{\pi}\kappa\label{E:FBG_Kappa}
\end{equation}
Examples of spectra with 1, 5 and 7 $\pi$-phase shifts are shown in Fig.~\ref{f:pi}. 

These $\pi$FBG gratings have found widespread uses in wavelength division multiplexing~\cite{agrawal_1994}, optical CDMA~\cite{Monga_2016}, distributed feedback lasers~\cite{westbrook_2011}, and sensing~\cite{huang_2011,ghosh_2023}. The simplest $\pi$FBG designs can achieve bandwidths of order 1 pm (100 MHz) and therefore can provide highly accurate wavelength selectivity.

In this work, the FBG refractive index modulation was written by a two-arm Sagnac interferometer. A 244~nm, 150~mW, UV laser (Sabre FReD) was launched on a uniform C-band phase mask such that the beam was split into two beams at the $\pm$~1 orders. These two beams are controlled by two acousto-optic modulators (AOMs) to control the amplitude and phase of the interference pattern. Prior to exposure, photosensitive fiber from CorActive Inc. was ‘soaked' with hydrogen gas at high pressure and temperature to increase the fiber's photo-sensitivity. The fiber was then exposed to the UV laser interference pattern, creating the refractive index modulation. The fabricated FBG devices were then annealed at 300~$^\circ$C for 10 minutes. This writing system is described in greater detail in Ref.~\cite{edvell_2014}, with a similar system reported in Ref.~\cite{bland-hawthorn_2016}.

\section{Results\label{RES}}
\subsection{Initial characterization}
Low resolution, low dynamic range characterisation of the fabricated $\pi$FBG devices was performed using a commercial Optical Spectrum Analyser (Advantest OSA Q8384). Although the dynamic range was limited to 60~dB dynamic range, the measurements confirmed the device center wavelength and that the measured spectrum was consistent with the design for attenuation less than 60~dB. Across five 7$\pi$-FBG devices, an average deviation of 25.8~pm from the target wavelength of 1536.3~nm was observed. Further characterisation was performed with an optical vector analyser (LUNA OVA 5000) to  measure the pass band insertion loss. The average insertion loss due to the 7$\pi$FBG devices, each with an expected rejection of over 100~dB, was 1.09$\pm$0.17~dB.

\subsection{High Dynamic Range Measurement\label{RES:SH}}
There are two significant challenges in measuring the filter performance over the entire filter dynamic range, which exceeds 12 orders of magnitude. First, measurements can easily be limited by the detection system dynamic range. Second, any source light outside the 40~GHz notch bandwidth, such as the broadband amplified spontaneous emission (ASE) background of the laser can contribute to a noise floor and limit the dynamic range. We used two measurement schemes to account for these difficulties: (i) a highly frequency-selective superheterodyne measurement discussed below, and (ii) a detector limited intensity measurement with heavy pre-filtering of the input laser discussed in the Supplementary Information. %Section~\ref{???}. %using high dynamic range detectors and a large amount of pre-filters.

%\subsection{Super-Heterodyne Measurement}
\begin{figure}[ht!]
\centering
\includegraphics[width=\linewidth]{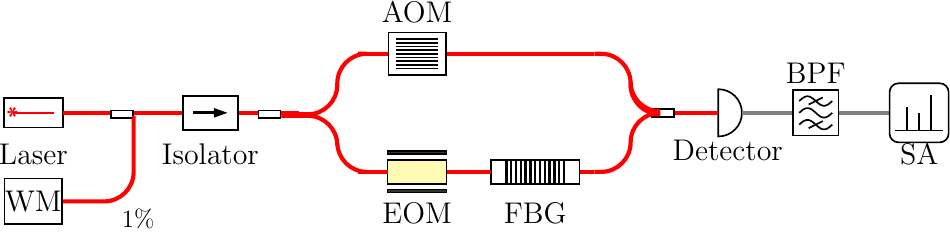}
\caption{Optical setup for superheterodyne measurements. A tunable laser is swept over the spectral region of interest and the frequency is monitored using a wavemeter (WM). An isolator prevents back reflections into the laser. The light is then split into two paths, one containing an acousto-optic modulator (AOM) and one containing an electro-optic modulator (EOM) and the FBG device under test. The light from these paths are recombined and measured using a high bandwidth detector. The output of the detector is filtered to a single beat frequency, which is monitored on the spectrum analyser (SA).}
%Insert Beamsplitter combination once finalised
\label{f:HeterodyneSetup}
\end{figure}
The primary method used to remove the ASE background of our laser source was a superheterodyne setup shown in \figref{f:HeterodyneSetup}. This setup used the beat signal between the +1 order sidebands of an Acousto-Optic Modulator (AOM) and an Electro-Optic Modulator (EOM). By selecting the operating frequencies of the modulators to be well-separated from any other sidebands, we isolated the amplitude change at the desired intermediate beat frequency with high spectral resolution. Two modulators were needed to avoid any leakage of the fundamental frequency through either modulator limiting the dynamic range. A detailed account of this measurement process is given in the Supplementary Information.

By sweeping the laser (Toptica CTL 1500) frequency we generated a spectrogram around the intermediate beat frequency. The laser frequency was measured by a wavemeter (HighFinesse WS6-200), the recombined light was detected on a low noise photo-receiver (Newport 1611) and the resulting intermediate beat frequency power was measured with a spectrum analyser (Fieldfox N9918A). We then determined the transmission through the FBG device under test at each laser frequency by taking the integral of the intermediate beat signal peak region above -3~dB relative to the signal peak. This was then normalised to dBm by a fit of the peak spectrum with an amplitude of 1~mW. The shape of the peak region is dominated by the windowing function of the spectrum analyser resulting in a Gaussian, with an amplitude weighted average FWHM of $1.054 \pm 0.008$~Hz. A full account of this procedure is included in the Supplementary Information.

As shown in \figref{f:ben1}, there is very close adherence to the design spectrum, modelled using coupled mode theory~\cite{erdogan_1997,kashyap_1999}, across the full filter spectrum. There are, however, some discrepancies between the design and fabricated device spectrum around the peak of the pass band in \figref{f:ben1} a). As discussed in Section~\ref{DIS}:\ref{DIS:Manufacture}, this is largely due to non-ideal fabrication resulting in the splitting of the pass-band peak. There is also a systematic contribution from measurement limitations not included in the error bars, whereby the scanning resolution and the high gradient of the filter combine to produce amplitude noise due to either laser frequency fluctuations or temperature-induced filter frequency shifts. Measurement of the observed pass bandwidth was complicated by a splitting of the central peak into a number of smaller peaks, each with a highly polarization dependant strength. When optimising the polarization to maximise transmission, the widest peak had a bandwidth of 290$\pm$30~MHz. From the model we expect a 3~dB bandwidth of the single, combined peak to be $2.184\pm0.005$~GHz. Details of the measurement of this splitting are provided in the Supplementary Information. The model was also used to extract a variety of parameters such as the slope of the filter and its bandwidth due to the limited number of points available in these regions from the experimental data. The average slope in the 0 to -30~dB region was -89.0$\pm$0.7~dB/GHz, significantly higher than those previously reported~\cite{deepa_2020}. Measurement of the slope is discussed further in the Supplementary Information.

%%%%%%%%%%%%%%%%%%%%%%%%%%%%%%%%%%%%%%%%%%%%%%%%%%%%%%%%%%%%%%%%%%%
\begin{figure}[ht!]
\centering
%\fbox{\inputpgf{7pi_heterodyne.pgf}{\linewidth}}
%\resizebox{\linewidth}{!}{\import{./figures}{7pi_heterodyne.pgf}}
\includegraphics[width=\linewidth]{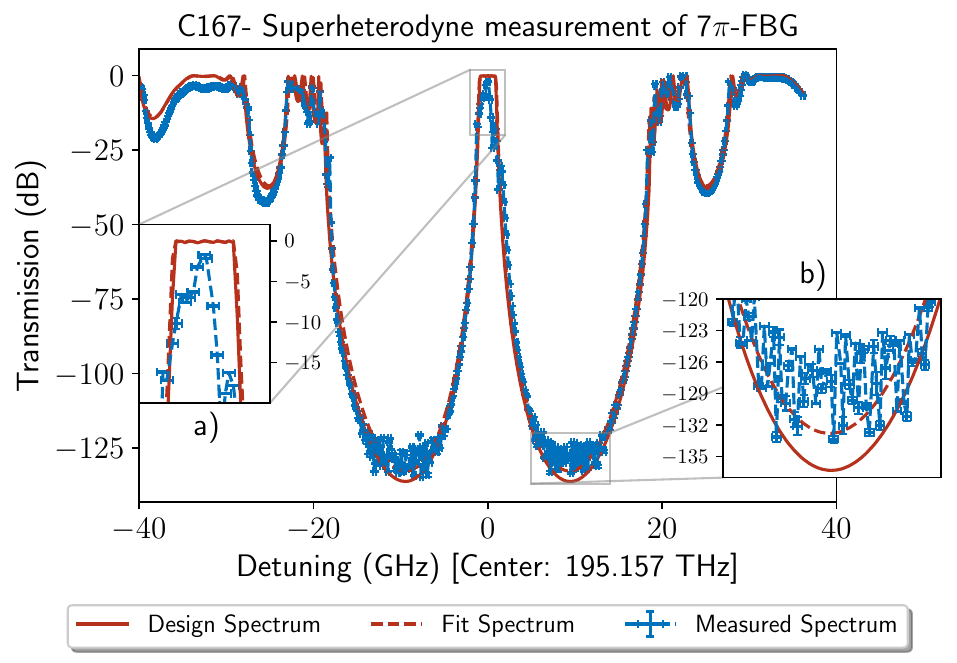}
\caption{Transmission spectrum of $7\pi$-FBG measured with superheterodyne detection. Two modelled spectra are included: the solid with the targeted $\kappa= 4.6$~cm$^{-1}$, and the dashed with the fitted achieved $\kappa = 4.5$~cm$^{-1}$. Inset a) shows the difference from the model in the pass band, as discussed in the text. Inset b) shows oscillation in the high suppression region, due to a low SNR.}
\label{f:ben1}
\end{figure}
%%%%%%%%%%%%%%%%%%%%%%%%%%%%%%%%%%%%%%%%%%%%%%%%%%%%%%%%%%%%%%%%%%%

In \figref{f:ben1} b) we see oscillations in the transmission at the highest suppression point of the filter as we approach the noise floor of the measurement setup. Accounting for this oscillation we see an average dynamic range of 127.6$\pm$0.3~dB, a maximum of  134.87$\pm$0.04~dB and a minimum of 122.2$\pm$0.3~dB. From the fitted spectrum we would expect an attenuation of -114.87~dB at a detuning of 5~GHz, from the experimental data we find an attenuation of -115.76$\pm$ 0.05~dB at 5~GHz, based on an average of the 5 nearest points weighted by the inverse of their detuning from 5~GHz. This exceeds the suppression requirement defined in the introduction for the quantum transduction application in a single device.

\section{Discussion\label{DIS}}
\subsection{Manufacturing Tolerances and Filter Performance\label{DIS:Manufacture}}
An important factor in the performance of the filters is how reliably they can be made to meet the design requirements. Across 8 devices of various suppression depths each targeting a central wavelength of 1536.3~nm we saw a standard deviation of 102~pm ($\approx 13$~GHz). These small detunings can be corrected by tuning with strain or temperature, with the measured sensitivities $(-153.1\pm0.6)$~MHz/$\mu\varepsilon$ and $(1.32\pm0.02)$~GHz/K, respectively (see Supplementary Information). For all devices in this fabrication run, we also observe splitting of the pass band into multiple peaks as shown in \figref{f:ben1}. Such deformations could be reproduced by modelling the filters with perturbations to the spacing and phase of the $\pi$-phase shifts (see Supplementary Information for further analysis). The consequence is a narrower pass band bandwidth, with the largest peak having a bandwidth of 446.9$\pm$0.2~MHz. The transmission at each peak frequency is polarization dependent due to both the inherent and stress-induced birefringence of the filter structure~\cite{Leblanc_1999,hnatovsky_2018}, requiring that the signal polarization be optimized to minimize loss in the wider peaks.

Another key performance metric is the device insertion loss, which is not yet optimized. For the main $7\pi$-FBG device measured in this work (C167), the connectorized device insertion loss was 1.619$\pm$0.002~dB at a detuning of $>$30~nm from the main filter structure, with an additional 0.66$\pm$0.04~dB of loss at the peak of the pass band (when polarization is optimized). Thus, the connectorized device insertion loss in the narrow passband was 2.28$\pm$0.04~dB. While this is comparable to filters used in transduction experiments~\cite{rochman_2023}, opportunities exist to significantly reduce this loss. The dominant loss mechanism is likely to be the  non-optimal splice between two fiber types with different mode field diameters (MFDs): SMF-28 (MFD: 10.4$\pm$0.5) fiber used for the connectors and CorActive (MFD: 6.6$\pm$0.6) photosensitive fiber containing the $7\pi$-FBG. The minimum loss for an ideal splice between these two fiber types is 0.9$\pm$0.8~dB~\cite{Ghatak_1998}. Writing FBG devices directly in SMF-28 fiber should significantly reduce insertion loss. For example, the insertion loss of conventional FBG devices written in SMF-28 fiber and connectorized demonstrated insertion losses comparable to the expected connector loss of 0.3~dB (see Supplementary Information for details).

\subsection{High Dynamic Range Measurements}
The superheterodyne measurement is well-suited to benchmark the filter performance against the intended design because it samples only coherent signals in a very narrow bandwidth (sub 10~Hz) around the intermediate frequency of interest. Thus, the noise rejection is extremely high. It does not, however, provide information about incoherent signals at any frequency or coherent processes that scatter photons into frequency modes outside this bandwidth. 

As an initial probe of the dynamic range limit imposed by coherent inelastic scattering processes such as Raman scattering~\cite{li_2022,patel_2012_1} and Brillouin Scattering~\cite{zadok_2022}, we repeated the superheterodyne measurement with a bandwidth of 1.5~GHz (limited by the photodetector bandwidth). At this increased bandwidth there is a significant reduction in the dynamic range (60~dB) and resolution (300~kHz). This measurement did not identify signals other than additional peaks arising from the beating between higher order modes of the AOM and EOM. We would expect the peak of Raman scattered light to be frequency shifted by of the order of 10~THz\cite{ribeiro_2008,li_2022}, outside of our ability to probe with the superheterodyne measurement. Given the 1~m length of the filter fiber and an input power of 0~dBm, we estimate the noise floor due to Raman scattering to be <-90~dBm~\cite{patel_2012_1}. We expect that Raman scattering will be limit on the dynamic range with a significantly smaller noise floor contribution from weak forward scattering from spontaneous Brillouin interactions.

We also performed an intensity measurement of the filter spectrum as a more informative lower bound on the dynamic range (see Supplementary Information for details). Here the challenge was to reduce the measurement limits imposed by the laser amplified spontaneous emission (ASE) noise, the laser Lorentzian linewidth and the photodetector noise. We pre-filtered the broad (1460-1570~nm) ASE background and Lorentzian wings of our probe laser using two FBG pre-filters to reduce the noise floor below the limit imposed by a commercial power meter (Thorlabs S154C). We observe a dynamic range limit of 75~dB, which is consistent with the probe power and the noise floor of the power meter. Further intensity measurements using more sensitive detectors are required to deconvolute the contributions from the probe laser, detector and fiber scattering processes. Optical pre-filtering, polarization control, strain control, and reducing coupling to cladding modes will all be important in measuring the ultimate dynamic range in such an intensity measurement.

\subsection{Broader applications of $\pi$FBGs}
In this work we have focused on the use of $\pi$FBG filters to suppress a strong classical pump beam, for microwave (5~GHz) to optical (telecom band) transduction. The design of these filters can be improved for lower insertion loss and larger attenuation at the pump frequency. The maximum attenuation of the $7\pi$FBGs studied in this work is $>$5~GHz detuned from the center of the pass band, whereas a 3$\pi$FBG device with a similar $\kappa =4$~cm$^{-1}$ would have a maximum attenuation $>$130~dB at 5.4~GHz. While increasing the pump suppression at 5~GHz, one trade off is the significantly narrower pass band at $\approx$50~MHz. In addition to optimizing the filter spectrum, the $\pi$FBG filters can also be designed for other center wavelengths relevant to transducer technology in other near-infrared windows~\cite{Yu_2024,kabakova_2009}. It is feasible to achieve the same filter performance as achieved in this work at other center wavelengths but requires verification given it is dependent on the susceptibility of the specific fiber materials to refractive index changes.

Filters incorporating a steep pass band within a deep notch are more broadly applicable to act as noise filters for a variety of single photon and entanglement generation sources~\cite{aharonovich_2016}, particularly if they can be realized at a wider range of centre frequencies. Examples of applicable technologies include single photon generation in quantum dots~\cite{coste_2023,senellart_2017} and emitters in diamond~~\cite{lee_2011}. The adaptability of FBG filter designs also makes them a candidate filter for quantum memory technologies~\cite{campbell_2016}. Particularly schemes similar to electromagnetically induced transparency (EIT)~\cite{hsiao_2018_1} require a single-photon output signal to be separated from a classical control beam with a small frequency separation. Each of the mentioned technologies will have their own requirements for the frequency detuning of the pump and signal, the bandwidth of the pass band, and the bandwidth and depth of the attenuating notch. While not all of these parameters can be tuned independently, the large gradients, high suppression and low insertion loss achieved in this work provide a foundation for exploring solutions for other applications both quantum and classical.

\section{Conclusion}
This work demonstrates the potential for fiber Bragg grating filters to be low loss components for hybrid microwave-optical quantum technologies to separate single photon signals from strong classical pump modes. We have designed and fabricated a device targeting an attenuation of over 100~dB at a detuning of 5~GHz from the filter's pass band center wavelength of 1536.3~nm. Our results demonstrate a high dynamic range measurement of these devices showing they closely match the design across the entire spectrum and achieve (-118.0$\pm$ 2.0)~dB of attenuation at 5~GHz. The maximum attenuation observed was (134.67$\pm$ 0.05)~dB with a pre-connector insertion loss of (1.09$\pm$0.17)~dB. Our results indicate that we have not reached the fundamental limit of either the dynamic range or insertion loss, and future work will improve device and measurement design to investigate the ultimate performance bounds. Beyond the applications in quantum transduction, our work highlights further opportunities for FBG technology to be enabling components in a wide range of precision photonic technologies.

%\clearpage

\begin{backmatter}
\bmsection{Acknowledgments} 
We are thankful to the Quantum Control Lab, Quantum Nanoscience Lab, and the Eggleton Group at the University of Sydney for equipment loans without which these measurements would not have been possible.

This work was performed in part at the Advanced fibre Bragg grating facility part of the Core Research Facility at the University of Sydney and the NSW node of the NCRIS-enabled Australian National Fabrication Facility (ANFF). 

This research was supported by the Australian Research Council Centre of Excellence for Engineered Quantum Systems (EQUS, CE170100009), LIEF grant (LE160100191) and Laureate Fellowship (FL140100278). BRF was supported by the Sydney Quantum Academy, Sydney, NSW, Australia. JGB acknowledges the support from the USYD CRF User Access Scheme, USYD SOAR prize and the Selby Scientific Foundation.

This material is also based upon work supported by the Air Force Office of Scientific Research under award number FA9550-21-1-0055. Any opinions, findings, and conclusions or recommendations expressed in this material are those of the authors and do not necessarily reflect the views of the United States Air Force

%\vskip 1ex

\bmsection{Disclosures} The authors declare no conflicts of interest.

\bmsection{Data Availability} Data underlying the results presented in this paper are not publicly available at this time but may be obtained from the authors upon request.

%\section{References}

\bigskip
% Bibliography
\bibliography{main}

% Full bibliography added automatically for Optics Letters submissions; the following line will simply be ignored if submitting to other journals.
% Note that this extra page will not count against page length
\bibliographyfullrefs{main}

\end{backmatter}
\end{document}